\documentstyle[pre,aps,tabularx]{revtex}

\begin{document}

\draft

\title{Scarred Patterns in Surface Waves}

\author{A. Kudrolli,$^{1,2,\dagger}$ Mathew C. Abraham,$^{1}$  and J. P.
Gollub$^{1,3,*}$ }

\address{$^{1}$Department of Physics, Haverford College, Haverford, PA
19041\\
$^{2}$Department of Physics, Clark University, Worcester, MA 01610 \\
$^{3}$Physics Department, University of Pennsylvania, Philadelphia, PA 19104}

\date{\today} \maketitle

\begin{abstract}

Surface wave patterns are investigated experimentally in a system
geometry that has become a paradigm of quantum chaos: the stadium
billiard.  Linear  waves in bounded geometries for which classical
ray trajectories are  chaotic are known to give rise to scarred
patterns.  Here, we utilize  parametrically forced surface waves
(Faraday waves), which become progressively nonlinear beyond the wave
instability threshold, to investigate the subtle interplay between
boundaries and nonlinearity. Only a subset (three main types) of  the
computed linear modes of the stadium are observed in a systematic scan.
These correspond to modes in which the wave amplitudes are strongly
enhanced along paths corresponding to certain periodic ray orbits. Many
other modes are found to be suppressed, in general agreement with a
prediction by Agam and Altshuler based on boundary dissipation and the
Lyapunov exponent of the associated orbit. Spatially asymmetric or
disordered  (but time-independent) patterns are also found even near
onset.  As the driving acceleration is increased, the time-independent
scarred patterns persist, but in some cases transitions between modes
are noted.  The onset of spatiotemporal chaos at higher forcing
amplitude often involves a nonperiodic oscillation between spatially
ordered and  disordered states.  We characterize this phenomenon using
the  concept of pattern entropy.  The rate of change of the patterns is
found to be reduced as the state passes temporarily near the ordered
configurations of lower entropy.  We also report complex but highly
symmetric (time-independent) patterns far above onset in the  regime
that is normally chaotic.

\end{abstract}



\section{Introduction}
Parametrically forced surface waves arising as a result of the Faraday
instability have provided an excellent opportunity to study nonlinear
pattern formation.  One of the special features of this system is that
the system size relative to the basic correlation length can be varied
so that both the large aspect ratio and small aspect ratio limits can
be explored.  At large aspect ratio, all of the  classic  ordered
patterns have been found, including stripes, hexagons, and squares;
additional exotic structures such as quasicrystalline and superlattice
patterns have also been found, as well as secondary instabilities
giving rise to spatiotemporal chaos.  Extensive references can
be found in~\cite{kudrolli96,binks97,kudrolli98}.

The case of small aspect ratio has also been studied in rectangular
and circular containers.  Typically the wave patterns found near onset
are either normal modes of the container or symmetrical combinations
of these modes~\cite{douady88}.  For example,  in the circular case
the normal modes are Bessel functions of the radius multiplied by
sinusoidal functions of the azimuthal angle.
The effects of container shape can be either a nuisance or a
benefit depending on one's point of view.  One example of the
usefulness of considering container geometry is the study by Lane
et al.~\cite{crawford93} in which the conceptual differences
between square symmetry and square geometry were elucidated.  On
the other hand, finite size effects  impede  efforts to
utilize amplitude equations to describe the wave dynamics.

The influence of the container shape is also a fundamental issue in
the field of quantum chaos.  There is known to be a close
correspondence between certain finite quantum systems (or
analogous  systems supporting  classical waves) and their particle (or ray-optic)
counterparts~\cite{gutzwiller91,heller93}.  Of particular interest
are non-integrable quantum systems with classical counterparts that
are chaotic, such as the billiard formed from two semicircles
separated by two straight edges.  For  almost all  initial conditions, a
particle launched inside such a billiard will exhibit sensitive
dependence on initial conditions - the hallmark of chaos.
Experimental, numerical and theoretical studies have shown that the
statistical behavior of the wavefunctions of the quantum or wave
version of this system is distinctly different from the
behavior  for ``integrable" geometries such as a square or
circle~\cite{kudrolli95,bohigas84,andreev96}.  Most notably,
regions of high  amplitude in the wavefunctions - called scars -
are found along some of the deleted word  periodic orbits of the
classical  counterpart~\cite{heller84}.

Effects of this type were explored to a limited extent using
parametrically forced surface waves by Bl\"umel et
al.~\cite{blumel92}. Their experiment utilized water as a working
fluid and high frequency excitation.  They reported observations of
``scarlets", that are ridge like structures consistent with a random
superposition of plane waves~\cite{connor87} (but are not located
along periodic orbits.) On the other hand, no clear evidence for the
simpler ``scarred" wavefunctions was given. The possible effects of
hydrodynamic nonlinearity on the utility of the ray optics approach has
also not been discussed.  Nonlinearity is in principle important, since
even infinitesimally above the onset of instability, saturation of the
wave amplitude is produced by nonlinear effects.

In this paper we first present observations of the spatial modes
of Faraday waves in a finite non-integrable geometry, close to onset
where nonlinearity is as weak as possible, and the waves might be
usefully described by quasi-classical (ray optics) methods.  A low
viscosity fluid in a stadium shaped container is used for this
purpose.  Scarred patterns that resemble the computed
eigenfunctions of the stadium geometry are clearly evident, but
some of the linear eigenfunctions are apparently suppressed.  The
relative suppression of certain modes has been plausibly explained
by Agam and Altshuler~\cite{agam99} in terms of higher  dissipation
rates for those modes near the boundaries in comparison with  the
modes that are observed.

We then consider the evolution of the wave patterns as the
degree of nonlinearity is increased.  Transitions between modes are
found at some driving frequencies, along with a general increase in
spatial complexity.  The scars that are characteristic of the
linear eigenmodes are often evident substantially above onset.
Finally, we consider the development of spatiotemporal chaos (STC)
in  the stadium geometry.  Though the onset of STC is strongly
dependent on the excitation frequency, the boundaries continue to
play a large role, leading for example to coherent oscillations
between symmetric and asymmetric states, a phenomenon that we study
using the concept of pattern entropy.

\section{Theoretical background}
A fluid layer with a free surface is subjected to an oscillatory
vertical acceleration of amplitude $a$.  The surface is flat until a
critical acceleration $a_c$ is reached, at which point the surface
becomes unstable and standing wave patterns are observed that
oscillate at half the driving frequency.  The threshold acceleration
depends on the frequency and the viscosity of the fluid.  It is
convenient to define  a dimensionless driving parameter $\epsilon =
(a - a_c)/ a_c$ that measures the  departure from onset and hence the
degree of nonlinearity. The  patterns are time-independent for a
range of positive
$\epsilon$ but eventually  a secondary instability gives rise to
spatiotemporal chaos.  In this  section, we briefly discuss the linear
inviscid theory, the effect of  viscosity, and the role of
nonlinearity, as they pertain to the  present investigation.

The linear stability theory for Faraday waves was first developed by
Benjamin and Ursell~\cite{benjamin54}. We summarize it here
because the quantum/classical correspondence occurs for linear
waves.  They started from the Euler  (inviscid) equation of motion and
the continuity equation for an ideal  fluid with a free surface in an
oscillating gravitational field, and  simplified the equations by
retaining only the linear terms  appropriate for small amplitude waves.
The surface deformation  $h({\bf x},t)$ as a function of  spatial coordinate 
${\bf x}$ and time $t$ may be written as a superposition of  normal modes
$\psi_i({\bf x})$ with coefficients
$A_i(t)$:

\begin{equation}
h({\bf x},t) = \sum_i A_i(t) \psi_i({\bf x}) \,\,\, ,
\end{equation}
where $\psi_i({\bf x})$ is a complete orthogonal set of eigenfunctions of
the  Helmholtz equation,

\begin{equation}
(\nabla^2 + k_i^2) \psi_i({\bf x}) = 0	.
\end{equation}

The sidewall boundary condition is imposed by setting the normal
 component of the  velocity of the fluid at the wall to zero.  This leads to a
quantization condition on $k_i$ (the wavenumber).  In addition, $k_i$
satisfies the dispersion relation which relates the frequency of
oscillation $\omega$ of the fluid to the wavenumber:

\begin{equation}
\omega^2 = \tanh(k_i d) ( \frac{\Gamma}{\rho} k_i^3 + g k_i ) \,\,  ,
\end{equation}
where, $\rho$ is the fluid density, $\Gamma$ is the surface tension,
$d$ is the mean  fluid depth, and $g$ is the gravitational
acceleration.  In our  experiment, the wave number is sufficiently
large so that the surface  tension term is much greater than the
gravity term.  The hyperbolic  tangent factor is close to unity since
$k_i d$ is large.

The time dependent amplitudes $A_i$ of these normal modes satisfy the
Mathieu equation:

\begin{equation}
\frac{d^2 A_i}{dt^2} + k_i \tanh k_i h ( k_i^2 \frac{\Gamma}{\rho} + g -
a \cos(\omega t)) A_i = 0  \,\,\,.
\end{equation}
An instability occurs and the amplitude $A_i$ grows exponentially
in time when the eigenvalue is in a band (known as the stability
tongue) such that  the frequency of oscillation of the fluid is half
the driving  frequency.  The instability occurs at arbitrarily small
driving  amplitude in the absence of viscosity.

Damping, which is provided by a number of distinct mechanisms in
addition to bulk viscosity, can be included by means of a
phenomenological linear damping term as reviewed in
Ref.\cite{cross93}.  Though treating damping in this way may not be
fully  adequate, the main effect is to reduce the width of the
stability  tongue in parameter space and raise the critical threshold
to a finite  amplitude.  A proper theoretical treatment of
instability in the viscous case has been given  in Ref.~\cite{kumar94},
where the shapes of the computed stability boundaries were presented.

If the acceleration $a$ is slightly higher than $a_c$, all modes in a
band  (${\bf k}-\Delta {\bf k}/2, {\bf k} +\Delta {\bf k}/2$) are
 accessible and can be excited.  The wavenumber width $\Delta k$ of the
stability band for small $\epsilon$ has been estimated
\cite{edwards94,kumar95} to be:

\begin{equation}
\Delta k = 8 \sqrt{2}\rho \nu \omega \sqrt{\epsilon}/3\Gamma
\,\,\, ,
\end{equation}
where $\nu$ is the kinematic viscosity of the fluid.  For a suitable
choice of $\epsilon$, $\rho$, and $\omega$, and assuming no
interaction  between modes, one then expects to find either single
mode patterns or  superpositions of a few modes whose thresholds lie
 in the window
$({\bf k}-\Delta {\bf k}/2, {\bf k}+\Delta {\bf k}/2)$.

The cumulative number of eigenvalues of the Helmholtz equation $N(k)$
is  related to the geometry and is given by:

\begin{equation}
N(k) \cong \frac{S}{4\pi} k^2  \mp  \frac{L}{4 \pi} k \,\, ,
\end{equation}
where $S$ is the area, $L$ is the perimeter of the stadium, and the negative or positive sign corresponds to Dirichlet or Neumann boundary condition respectively
\cite{gutzwiller91}. At high $k$, the perimeter term is negligible
compared to the area term. Taking $\epsilon = 0.01$ and using Eqs. 5,
6 with an area $S$ that is appropriate  to the experiments reported
here, one can estimate that the typical number of accessible  modes
is about 8 for a driving frequency of 70 Hz.

What are the effects of nonlinearity?  A nonlinear theory that
describes regular Faraday wave patterns in large
containers rather well has been given by Zhang and
Vi\~nals~\cite{zhang96} and Chen and  Vi\~nals~\cite{chen97}. In this
theory, an evolution equation is  determined for the time derivative
of the amplitude of a typical  Fourier mode $B_1$ of the interfacial
deformation.  It may be expressed  in the form

\begin{equation}
\frac{dB_1}{dT} = \alpha B_1 - g_0 B_1^3 - \sum_{m \neq 1}
g(\theta_{m1}) B_m^2 B_1 \,\, ,
\end{equation}
where $T$ is a slow time variable, the linear term is due to the basic
instability, the cubic self-interaction term produces saturation of
the wave pattern, and the coupling terms to other modes (which depend
on their relative angle $\theta$) are also of cubic order.  The
constants  have been computed, and the ratio $g(\theta)/g_0$ is of order
unity and  independent of $\epsilon$.  This implies that coupling effects
between the  accessible modes may be substantial.  The theory was able to
explain the striking cascade of 2n-fold patterns discovered by Binks and
van de Water~\cite{binks97}.   It also explains
semi-quantitatively the appearance of striped, square and hexagonal
patterns observed in experiments using viscous fluids in large
containers~\cite{kudrolli96}. However, the amplitude equation  is
variational, and is only appropriate near onset.  It cannot  describe
nonuniform patterns, secondary instabilities, or  spatiotemporal chaos.
An earlier approach that allowed spatially  varying patterns was given by
Milner~\cite{milner91}.

The amplitude equations also ignore the effects of the boundary. For
slightly viscous fluids in small containers, a large fraction of the
dissipation occurs in the boundary layer and can in fact be the leading
cause of dissipation~\cite{agam99,mei73,martel98}.  In work
stimulated by the experiments reported here, Agam and
Altshuler~\cite{agam99} show that the dissipation near the boundary
depends strongly on the nature of the mode.

\section{Experimental Apparatus}

The apparatus is similar to that used by Gluckman {\em et al.} in
Ref.~\cite{gluckman95}.  Fig.~1 shows a schematic diagram of the
experimental setup.  The stadium shaped container made of Delrin has the
following dimensions: depth $d = 1.25$\,cm, radius of semicircles
$r= 3$\,cm, and length of straight edge $l = 4.5$\,cm.  The top and
bottom  plates of the container are made of glass to allow the
transmission of  light.  The fluid is silicone oil of kinematic viscosity
$0.02 \,{\rm cm^2s}^{-1}$, chosen for its stable surface tension and good
wetting  characteristics.  To minimize meniscus waves, a brim full
boundary  condition was prepared by machining a ledge in the boundary at
the  same height as the fluid.  Therefore the fluid meets the ledge at
$90^{0}$.   By maintaining the fluid under brim full conditions, the fluid
surface  is pinned to the ledge and boundary dissipation is reduced.
This situation has been modeled as a Dirichlet  boundary
condition ($\psi =0$)~\cite{douady90}. The container is  rigidly attached
to an electromagnetic shaker (Vibration Test Systems  Model 40C) and the
acceleration is measured with an accelerometer.   The apparatus is placed
within a temperature-controlled environment.   The driving frequency is
selected to be greater than 55~Hz to be in  the capillary wave limit, but
less than 75~Hz to prevent the density  of modes from becoming too high.

The patterns are imaged with shadowgraph techniques.  The specific
implementation is discussed in depth in~Ref.\cite{gluckman95}.
Light from an expanded and collimated incident beam is collected and
imaged onto a CCD (charge-coupled device) video camera via a large
collecting lens and the camera lens.  The resulting images can be
interpreted by considering which rays of light reach the CCD plane
after passing through the fluid.  The relatively small aperture of the
camera lens restricts the rays that reach the CCD. All the
rays that leave the fluid surface at an angle measured from the
normal that is greater than a critical angle (typically about
$10^{-2}$  radians) are blocked.  Since the critical angle is so
small, light is  collected only from the nearly horizontal regions of
the wave surface.   Therefore, the bright regions in an image
corresponds to local extrema  or antinodes of the wave pattern.
Images are averaged over one video  frame, 1/30~s, which is more than
a full cycle of the standing waves.   The imaging process is
nonlinear in the wave height, but a  quantitative model for the
measured intensity was presented and tested  in
Ref.~\cite{gluckman95}.

\section{Patterns Near Onset}

We made a survey of the time-independent wave patterns near
onset over the range 55 to 65~Hz by changing the frequency in 0.1~Hz
steps. In order to obtain
useful statistics for the surface wave patterns, a systematic procedure
was followed: for  each selected frequency, $a_c$ was first measured to
within 0.1\% and  then $\epsilon$ was raised to 0.01, the smallest value
that could be maintained  accurately. The threshold $a_c$ is
$3.1\,{\rm m\ s^{-2}}$ at $f = 60$ Hz and  increases weakly with frequency .

A sequence of images from the survey for
driving frequency $f$ between 60.1~Hz and 62.8~Hz, and with an
approximate spacing of 0.4~Hz (i.e. every fourth image), is shown in
Fig.~2.  This spacing is comparable to the experimentally observed
increment (0.3~Hz) typically required to obtain a distinctly different
pattern in this frequency range; it is greater then the computed mean
level spacing (about 0.1~Hz in this frequency range as  estimated from
Eqns.~5 and 6.) Most of the observed patterns show the reflection
symmetries of the
stadium.  Regions of large amplitude are often  located along lines that
would form  periodic  orbits of the classical  analog.  Since these regions
are similar to those found in other  numerical and experimental
investigations \cite{heller84,sridhar91}, we  refer to the patterns
containing such enhancements as scarred patterns.

We compare the observed patterns with numerically computed
eigenstates  of the Laplace operator for the stadium geometry,
obtained for  comparable mean wavenumber using an algorithm due to
Heller \cite{heller97}. Sample computed eigenstates (selected from a
large  number of distinct patterns) are shown in Fig.~3.  We find
that some  of the computed states resemble observed patterns.  On the
other hand,  a one-to-one correspondence for sequences of eigenstates
was  definitely not observed.  Furthermore, certain computed
eigenstates  that occur frequently  such as the ``whispering gallery" mode (Fig.~3d) were not observed in the full experimental frequency range.

Interestingly, most of the observed symmetric patterns resemble one of
three basic classes of eigenstates shown in Fig. 3(a-c).  For instance,
Figs. 2(a,d) resemble the bouncing ball eigenstate Fig. 3(a);  Figs.
2(g,h) are close to the longitudinal eigenstate of Fig. 3(b); and
Figs. 2(b,e) are a combination of the longitudinal and bowtie eigenstates
of Figs. 3(b,c).  It is noteworthy that among the observed patterns are
states such as Fig. 2(c) that do not have the reflection symmetries of the
stadium.  We refer to these as disordered  patterns.  Table I summarizes
the percentages of the onset patterns that were visually judged to
approximate particular computed scarred eigenstates in the frequency
range 55 to 65 Hz at $\epsilon = 0.01$. (Visual comparison was used
because automated pattern recognition, which we attempted, was not
sufficiently reliable.)

Since the discovery of scars, there have been a number of theoretical
attempts to obtain a quantitative measure for
scarring~\cite{bogomolny88,agam94,kaplan99} based on eigenstate overlap,
Wigner function overlap, and inverse participation ratios for the
amplitudes in the vicinity of the scars. To utilize such measures
experimentally, the local wave amplitude is required with high accuracy.
The shadowgraph technique used here is quantitative but nonlinear
~\cite{gluckman95} and does not provide this information. Development of a
quantitative experimental measure of scarring has proven to be difficult
even for linear probes.

We use the concept of ``pattern entropy" as a tool to classify
the patterns.  Egolf, Melnikov, and Bodenschatz~\cite{bodenschatz97} have
applied this concept successfully to measure the complexity of patterns
observed in Rayleigh-B\'enard convection. The pattern entropy is
calculated from the power spectrum of the pattern.  If $P({\bf k})$
is the normalized two dimensional power spectrum of  the pattern at time
t, then the pattern entropy $E(t)$ is defined as:

\begin{equation}
E(t) = - \sum_{{\bf k}} P({\bf k}) \ln(P({\bf k})) \,\,  .
\end{equation}
Here $E(t)$ measures the spectral complexity of a pattern. If
the image consists of just one Fourier mode of amplitude unity, then $E =
0$; otherwise $E > 0$.  To minimize the effects of experimental noise, we
sum contributions only in a band of wavenumbers centered at  the mean
wavenumber of the pattern with a rage of $\pm 25 \%$. In Table II, the
approximate entropy ranges for the various types of patterns observed in
the range 55-65 Hz are given. Note that the patterns are not
distinguishable solely by their entropy, since some of the ranges
overlap. However, the pattern entropy can be useful in studies of time
dependence farther above onset, as we show in Sec. VI.

\section{Patterns beyond onset}

Here we examine the evolution of the wave patterns farther from onset,
where the interactions between different Fourier components of the waves
become increasingly nonlinear and the approximation of Eq.~(2) becomes
inapplicable.  The patterns were observed to be time-independent while
changing adiabatically with $\epsilon$ for $\epsilon< 0.3$.  On the
other hand, they become weakly time-dependent for $\epsilon
\ge 0.3$ at most frequencies.

The evolution toward time-dependence with increasing $\epsilon$
depends on the excitation frequency.   Three  examples of this evolution
are shown in  Figs. 4-6.  For some driving frequencies, the patterns
remain reflection-symmetric as $\epsilon$ is increased, but exhibit
transitions from one spatial mode to another prior to the onset of time
dependence, as in Fig. 5.  In these cases, the transition to spatial
disorder (asymmetry) tends to coincide with the onset of spatiotemporal
chaos (STC).

It is important to note that as $\epsilon$ increases, the width
of  the stability tongue grows (see Eq.(5)): for example at $f =
74.1$\,Hz and $\epsilon = 0.2$ the number of accessible modes of the
container is  approximately 35.  Therefore, the observed mode switching
might be a combined effect of the increase in the number of accessible
modes and an increase in the degree of nonlinearity that
couples them.  It is remarkable that the container boundary
continues to influence the patterns even at $\epsilon=0.252$ (Fig.
5d), where nonlinearity clearly plays a major role.

The variability of the nonlinear development is evident from
examining the examples in Figs. 4-6.  In Fig. 4, there is a general
increase in complexity with $\epsilon$, but the dominant mode does not
change.  In Fig. 6, the near onset pattern is nearly obliterated even at
$\epsilon = 0.025$ by the growing complexity, and the pattern is
also distinctly asymmetric, while remaining time-independent.

In one instance, a complex but symmetric time-independent pattern was
observed at an unusually high driving amplitude of $\epsilon = 0.8$ at a
frequency of 65.0 Hz, in a regime where spatiotemporal chaos is usually
fully developed.  The image shown in Fig.~7 was averaged over 3000 images
taken over a period of 5 minutes  to test for time-dependence.  The lack
of blurring demonstrates its time-independence.

\section{Role of ordered states in the regime of spatiotemporal chaos}

For driving amplitudes just beyond the frequency-dependent onset of
spatiotemporal chaos, the time dependence  of the pattern is often
intermittent; the patterns appear to oscillate  between states that are
relatively ordered and states that are relatively disordered (see
the images in Fig.~8 and a corresponding web-based movie
~\cite{movie}).  The  power spectra of these typical patterns are also
shown in Fig.~8 and indicate the greater complexity of the disordered
case, where the power is distributed more uniformly on the ring
corresponding to the dominant wavenumber.  The time  dependence and
complexity of the patterns are  monitored using two quantitative
measures: (i) the rate of change of the pattern $R(t)$, and (ii) the
entropy $E(t)$ as defined in Eq.~(8).  The  rate of change $R(t)$ was
calculated by subtracting two consecutive images I(${\bf x},t +\Delta t$)
and I(${\bf x},t$) separated by a time interval  of $\Delta t = 0.36$\,
seconds, and calculating $R(t)$ according to the following  formula:

\begin{equation}
R(t) = c \sum (I({\bf x},t+\Delta t) -I({\bf x},t))^2	\,\,,
\end{equation}
where, ${\bf x}$ is the position, and $c$ is a constant scaling factor.

In Fig. 9(a), a section of the resulting quantity $R(t)$ is shown
as function of time for $f = 71.9$\,Hz and $\epsilon = 0.55$.  The
data has been smoothed by averaging  over four adjacent points.  At every
pronounced minimum of this  smoothed data we find that the corresponding
pattern is symmetric and  appears to have long range order.  At all other
times the pattern is  asymmetric and  disordered.  A graph of the
corresponding entropy $E(t)$ is shown in Fig.~9(b).  Examples of the
ordered (X) and disordered (Y) states are shown in Fig.~8, where the
location in time is indicated by  symbols X ($t = 124$\,s) and Y ($t =
136$\,s) respectively in Fig.~9.  At X, where the pattern entropy is low,
$R(t)$ is small, while at Y, where the  pattern entropy is high,
$R(t)$ is large.  This correlation between $R(t)$, and $E(t)$ holds true
for most of the other strong peaks and valleys.

For higher $\epsilon$ ($ > 0.8 $) the oscillations diminish in
strength and  uniform STC is observed. In this regime, following
Gluckman et  al.~\cite{gluckman95}, we obtained the time averaged
pattern after  adding 3000 instantaneous images obtained over five
minutes. An  example of an instantaneous pattern is shown in
Fig.~10(a).  The  resultant average pattern is shown in Fig.~10(b).
The average reveals considerable ordered structure including remnants
of the bouncing ball states.   This phenomenon is seen at
most frequencies and persists up to  $\epsilon \sim 1.6$.  Beyond this point, scars were visually absent  and the average patterns are locally parallel
to the boundary, as observed in circular or square patterns by Gluckman et
al.~\cite{gluckman95}.

\section{Discussion}

In this paper we have discussed the parametrically forced wave patterns
formed in a stadium-shaped container containing a low viscosity fluid, as
a function of driving frequency and amplitude.  The patterns near onset
($\epsilon=0.01$) were compared to a simple model consisting of
linearized equations that reduce to the Helmholtz equation with Dirichlet
boundary conditions (see Section II).  While a large proportion of
observed patterns resemble the numerically  computed eigenstates of the
stadium, many of the computed eigenstates  (for instance the whispering
gallery modes) were not observed in a scan with sufficient frequency
resolution to detect them. The observed  patterns may be broadly
classified into three categories: (a) bouncing  ball patterns, (b)
longitudinal patterns, and (c) bowtie patterns,  which have high
amplitudes near corresponding periodic orbits.   In addition, a
significant number of disordered patterns (lacking in symmetry but
time-independent) were observed near onset.  Furthermore, the observed
mode spacing ( $\sim 0.3$\,Hz) is somewhat greater than the mean
eigenvalue separation  implied by Eq.~(6) ($\sim 0.1$\,Hz).  These
observations imply that the simplest model is inadequate even close to
onset.

Recently, Agam and Altshuler have offered an explanation for the
selection of  modes at onset~\cite{agam99} by considering the
stability of the periodic orbits corresponding to the scars. They
argue that a threshold for excitation of a particular scarred pattern
is given by:

\begin{equation}
h > \gamma_{b} + \gamma_{p} + \lambda/2   \,\,  ,
\label{criteria}
\end{equation}
where $h$ is proportional to the rate that energy is pumped into the
system (i.e. the driving amplitude), $\gamma_{b}$ is the
dissipation rate in the bulk of the fluid, $\gamma_{p}$ is the
dissipation near the perimeter, and $\lambda$ is the Lyapunov
exponent of the ray orbit that predominantly scars the pattern. The
bulk dissipation $\gamma_{b} = \nu k^2$ is the same
for all the patterns and corresponds to approximately 2 sec$^{-1}$ in the
frequency window used in the experiments. Therefore the appearance
of a scarred pattern depends on a combination of the two remaining
factors which are orbit-dependent. Stability of a pattern is
favored both by a small Lyapunov exponent of the associated
scarring orbit, and by small perimeter dissipation $\gamma_{p}$.

In the limit of high wavenumber $k$ relevant to our experiments,
Agam and Altshuler derive an expression for the  damping rate of
scars due to boundary effects:

\begin{equation}
\gamma_p= \frac{\sqrt{\omega \nu/2}}{L} \sum_i \frac{(1-
\cos^2(\phi_i))}{\cos(\phi_i)}
\label{criteria2}
\end{equation}
where $\omega$ is the angular frequency, $\nu$ is the viscosity, $L$
is the  length of the periodic orbit, and the sum is over all the
collision points of the orbit with the boundary, $\phi_i$ being the angle
between the orbit and  a line perpendicular to the boundary at the
collision point. The parameter $\gamma_p$ and the Lyapunov exponent
have been calculated for most of the shortest periodic orbits and a long
ergodic orbit in Ref.~\cite{agam99}. The perimeter dissipation
$\gamma_p$, which can be either smaller or larger than $\gamma_b$, varies
between 0.2 sec$^{-1}$ and 3.0 sec$^{-1}$.  The extreme values correspond
to patterns scarred by the horizontal orbit and the ergodic orbit
respectively.

Most of the features observed in the experiments near onset appear to be
captured by Eq.~(\ref{criteria}).  The bouncing ball orbit is
prominent because the Lyapunov exponent is zero in that case, and the
longitudinal orbit occurs because of relatively low perimeter
dissipation. The whispering gallery orbits and others with angles
that come close to  $\pi/2$ have particularly large boundary dissipation
and are suppressed.  Eq.~\ref{criteria} implies that if one increases
the dissipation on the boundary so that it dominates, the longitudinal
orbit will be the last to survive. This is precisely what we observe in
the experiments when the level of the fluid is lowered, a change that
results in higher perimeter dissipation because of motion of the contact
line.

The theory just cited~\cite{agam99} is also able to account for the
observed tendency of one scarred pattern to suppress other nearby
eigenmodes through nonlinear interactions, as well as the existence of
some asymmetric patterns. At some driving frequencies the patterns
are observed to switch modes as $\epsilon$ is increased (Fig.~5). This
occurs especially when the  bowtie state is observed at onset, an
observation that may be related to the larger perimeter
dissipation and Lyapunov exponent of the bowtie mode.

At higher driving amplitude, additional nonlinear effects occur as
indicated by the growth in spatial complexity, and no adequate
theoretical treatment exists. However, the onset patterns are often
robust, persisting in the presence of increasing spatial complexity. The
boundaries remain influential even beyond the onset of time dependence.
At sufficiently high $\epsilon$ ($\sim 0.2$),  the onset patterns are no
longer visible, though they persist in the time-average.

Strong intermittency in the degree of order of the patterns is
observed in the regime of spatiotemporal chaos.  Furthermore, the rate of
change of the pattern $R(t)$ just above the STC onset is strongly
correlated with the order as characterized by the entropy $E(t)$.  The
more ordered patterns evolve more slowly in time, a striking observation
that remains to be explained. This tendency for ordered patterns to be
more stable may be related to the complex but highly symmetric
and time-independent pattern observed at atypically large
acceleration (Fig.~7).

\section{Acknowledgments} We thank Oded Agam and Boris Altshuler for
many useful discussions, and E.J. Heller for software to determine
eigenstates of the stadium.  Bruce Boyes provided technical help. This
work was supported by the National Science Foundation under Grant
DMR-9704301.  A. K. acknowledges a grant from the Alfred P.
Sloan  Foundation.

\begin{figure}
\caption{The Faraday wave and shadowgraph imaging apparatus. A stadium
shaped container is rigidly attached to an electromagnetic shaker,
which provides a sinusoidal oscillation of amplitude $a$.}
\end{figure}


\begin{figure}
\caption{Shadowgraph images of patterns observed in a stadium shaped
container filled with silicone oil as a function of driving frequency
($\epsilon = 0.01$; $\nu = 0.02$\,cm$^2$ s$^{-1}$).  The frequency
interval between displayed images is comparable to that required for the
pattern to change significantly.  (a) 60.1\,Hz; (b) 60.4\,Hz; (c)
60.8\,Hz; (d) 61.2\,Hz; (e) 61.6\,Hz; (f) 62.1\,Hz; (g) 62.4\,Hz; (h)
62.8\,Hz.  The patterns are strongly influenced by the shape of the
container.  Both symmetric scarred patterns and asymmetric patterns are
found (see text).}
\end{figure}


\begin{figure}
\caption{Selected eigenfunctions of the Helmholtz equation for a
stadium  geometry identical to that used in the experiments.
Wavenumbers are chosen  to be in the range explored experimentally.
(a) Bouncing ball mode;  (b) longitudinal mode; (c) bowtie mode; (d)
whispering gallery mode. A large  proportion of the experimental patterns
are similar to (a-c), which are scarred by periodic orbits (shown by
lines) of the  corresponding ray system.  However eigenstates such as the
whispering gallery mode (d) are not observed in the experiments.}
\end{figure}


\begin{figure}
\caption{Steady patterns as a function of driving amplitude
$\epsilon$ illustrate  the effect  of increasing nonlinearity at $f =
62.8$\,Hz. (a) $\epsilon = 0.01$;  (b) $\epsilon = 0.025$; (c)
$\epsilon = 0.125$; (d) $\epsilon = 0.249$. The  influence of the
scarred eigenfunction persists even in the presence of  strong
nonlinearity. Spatiotemporal chaos develops at higher $\epsilon$,
beyond the range shown here.}
\end{figure}


\begin{figure}
\caption{Same as  Fig.~4, except that $f = 71.4$\,Hz. (a) $\epsilon =
0.010$; (b) $\epsilon = 0.020$; (c) $\epsilon = 0.078$; (d)
$\epsilon = 0.252$.  In this case, the scars persist but the dominant
mode switches.  The ``bowtie" enhancement is lost and a ``bouncing ball"
enhancement appears. }
\end{figure}


\begin{figure}
\caption{Same as Fig.~4, except that $f = 74.7$~Hz. (a) $\epsilon =
0.010$;  (b) $\epsilon = 0.015$; (c) $\epsilon = 0.025$; (d)
$\epsilon = 0.058$ for $f = 74.7$\,Hz.  In this case, the pattern
becomes disordered at rather low $\epsilon$, but remains
time-independent.}
\end{figure}


\begin{figure}
\caption{A stationary symmetric pattern that is quite complex,
observed  at $\epsilon = 0.8$ ($f = 65$ Hz), a regime where
spatiotemporal chaos usually  predominates.  This image is an average
over 5 min; its sharpness demonstrates its stationarity.}
\end{figure}


\begin{figure}
\caption{The onset of spatiotemporal chaos involves an irregular
oscillation  between (a) relatively ordered and (b) disordered
patterns ($\epsilon =  0.55$ and $f = 71.9$\,Hz).  The corresponding
power spectra are shown in  (c) and (d) respectively. The greater
spectral isotropy in (d) contributes  to a higher pattern entropy.}
\end{figure}


\begin{figure}
\caption{(a) The rate of change $R(t)$ of the pattern as a function
of time,  as computed by differences between successive images. (b)
The pattern  entropy $E(t)$ as a function of time ($\epsilon = 0.55$,
and $f =  71.9$\,Hz). Both  functions oscillate in time and they
are strongly correlated; both are smaller for ordered patterns (e.g.
point X, shown in Fig.~8(a)) than for disordered patterns (e.g. point Y,
shown in Fig.~8(b)).  A movie corresponding to this figure is available
at the internet address in Ref. [33].}
\end{figure}


\begin{figure}
\caption{(a) Instantaneous and (b) time-averaged patterns in the
regime of  spatiotemporal chaos ($\epsilon = 1.2$, $f = 65$\,Hz).  We
find that scars  usually persist (except at very high $\epsilon$) and
that the dominant pattern  found at most frequencies is the one
shown.}
\end{figure}

\begin{table}
\begin{center}
\begin{tabular}{l c}
\hline
Class of Pattern\,\,\,\,  &	\,\,\,\, Percentage of Occurrence \\
\hline
\hline
Bouncing ball	& 27.2\%  $\pm$ 1.6\% \\
Longitudinal &	50.1\% $\pm$  3.3\% \\
Bowtie	& 13.2\%  $\pm$  2.5\% \\
Disordered &	13.2\% $\pm$ 2.5\% \\
\hline
\end{tabular}
\end{center}
\caption{ The percentages of patterns of various types observed
in a sample of 128 patterns near onset ($\epsilon = 0.01$), for
driving frequency between 55 and 65~Hz. A typical example of a bouncing
ball pattern is shown Fig. 2(a), a longitudinal pattern in Fig. 2(g),
a bowtie pattern in Fig. 2(e), and a disordered pattern in  Fig. 2(c).
Patterns with several components such as Fig. 2(b), which contains both
the longitudinal and bowtie modes, are counted in both categories.
Therefore the sum of the percentages is slightly over 100\%.}
\end{table}

\begin{table}
\begin{center}
\begin{tabular}{l c}
\hline
Class of pattern\,\,\,\,	& \,\,\,\, Approximate entropy ranges \\
\hline
\hline
Bouncing ball &	3.0 to 4.2 \\
Longitudinal &	4.2 to 5.6 \\
Bowtie	& 4.5 to 5.8 \\
Disordered &	5.8 to 6.4\\
\hline
\end{tabular}
\end{center}
\caption{Pattern spectral entropy, which is a measure of complexity
and is used in classifying the observed patterns. The patterns are
obtained in the driving frequency range 55 to 65 Hz, slightly above onset
($\epsilon = 0.01$).}
\end{table}


\end{document}